\documentclass[12pt]{iopart}

\usepackage{graphicx}% Include figure files

\begin{document}

\title{Large scale effects on the decay of rotating helical and non-helical turbulence}

\author{T Teitelbaum$^{1}$ and P D Mininni$^{1,2}$}

\address{$^1$ Departamento de F\'\i sica, Facultad de Ciencias Exactas y
         Naturales, Universidad de Buenos Aires and CONICET, Ciudad Universitaria, 1428
         Buenos Aires, Argentina. \\
             $^2$ NCAR, P.O. Box 3000, Boulder, Colorado 80307-3000, U.S.A.}
\ead{teitelbaum@df.uba.ar}

\begin{abstract}
Decaying three-dimensional (3D) turbulence is studied 
via direct numerical simulations (DNS) for an isotropic 
non-rotating flow and for rotating flows with and without 
helicity. We analyze the cases of moderate Rossby number 
and large Reynolds number focusing on the behavior of 
the energy spectrum at large scales and studying its 
effect on the time evolution of the energy and integral 
scales for $E(k)\sim k^4$ initial conditions. In the 
non-rotating case we observe the classical energy decay 
rate $t^{-10/7}$ and a growth of the integral length 
proportional to $t^{2/7}$ in agreement with the prediction 
obtained assuming conservation of the Loitsyanski integral. 
In the presence of rotation we observe a decoupling in 
the decay of the modes perpendicular to the rotation 
axis from the remaining 3D modes. These slow modes show 
a behavior similar to that found in two-dimensional (2D) 
turbulence whereas the 3D modes decay as in the isotropic 
case. We phenomenologically explain the decay considering 
integral conserved quantities that depend on the large scale 
anisotropic spectrum. The decoupling of modes is also 
observed for a flow with a net amount of helicity. In 
this case, the 3D modes decay as an isotropic fluid with 
a constant, constrained integral length, and the 2D modes 
decay as a constrained rotating fluid with maximum helicity.
\end{abstract}
\maketitle


\section{Introduction}

Controversy on the invariance of integral quantities in decaying turbulence has aroused 
during the last years \cite{Davidson 2004}. Integral invariants are required to build 
phenomenological theories to explain the decay of kinetic energy 
and the evolution of integral length scales. For isotropic and homogeneous turbulence, 
the conservation of the Loitsyanski integral $I$ for an 
initial spectrum $E(k\rightarrow0) \sim Ik^4$ was called to derive the classical 
energy decay rate $E \sim t^{-10/7}$ and a growth of the integral length $L$ proportional 
to $t^{2/7}$ \cite{Kolmogorov 1941}. In a similar fashion, for an initial 
spectrum $E(k\rightarrow0) \sim Sk^2$, the assumed conservation of the integral quantity $S$ 
associated to the conservation of linear momentum leads to $E \sim t^{-6/5}$ 
\cite{Saffman 1967}. In practice, theses quantities were shown to be only approximately 
conserved in closures \cite{Herring 2005} and in numerical simulations 
\cite{Ishida 2006, Davidson 2007} depending on the large scale spectrum of 
the initial conditions.

Less is known about the decay of turbulent flows in rotating reference frames.
These flows have been largely studied due to its relevance for 
scientific and engineering problems. Their importance has 
motivated numerous theoretical, experimental, and numerical works. Applications 
are broad, including areas as diverse as turbo machinery and rotor-craft, 
convective region of the sun and stars, large-scale flows in oceans, and convective 
scales in the atmosphere. 

It is well known that solid-body rotation inhibits the non-linear direct 
cascade of energy toward small scales reducing the dissipation 
rate of kinetic energy in comparison with non-rotating flows. This 
reduced dissipation has been observed in simulations \cite{Mansour 1992,Squires 1994}, 
experiments \cite{Morize 2005,Morize 2006}, and studied 
theoretically \cite{Cambon 2004}. 
An increase in the integral length parallel to the 
rotation axis with time has also been reported \cite{Bardina 1985, Bartello 1994} for 
these flows.

Resonant wave theory has been used to take into account the effect of rapid rotation 
in turbulence \cite{Greenspan 1968,Greenspan 1969,Waleffe 1993,Cambon 1997}. 
According to this approach, the energy is transferred from small to large 
scales by resonant triadic interactions of inertial waves. 
The theory also argues that the resonant interactions are responsible 
for driving the flow to a quasi-two-dimensional state. In resemblance with the 
classic Taylor-Proudman theorem for steady flows \cite{Greenspan 1968}, this 
result is often called the ``Dynamic Taylor-Proudman Theorem'' (see eg. \cite{Chen 2004}), 
and leads to the decoupling of slow modes which behave as 
an autonomous system of two-dimensional (2D) modes for 
the horizontal velocity components (perpendicular to the rotation axis) for strong rotation 
\cite{Waleffe 1993, Majda 1998}.

As a result of this reduced non-linear coupling and dissipation, for decaying flows in 
the laboratory different scaling laws were observed 
as the flow decays \cite{Morize 2005, Seiwert 2008}, from classical 
non-rotating values at small times changing to different power laws after a time of 
the order of $1/\Omega$. In \cite{Moisy 2009}, for example, an initial isotropic decay is observed, 
followed by a cross-over for $R_o \approx 0.25$ after which the energy decays slower 
($E(t) \sim t^{-3/5}$). Such a decay law was proposed by \cite{Squires 1994} 
based on the assumption of energy tranfer being governed by the linear time $\Omega^{-1}$. 
Strong correlation of the vertical flow leading to the growth and subsequent saturation of the integral 
length by vertical confinement was observed in \cite{Morize 2006, Bardina 1985, Moisy 2009}. 
This saturation was observed at a time proportional to $\Omega^{-5/7}$ in \cite{Morize 2006}. 
In \cite{Staplehurst 2008} large-scale columnar-structure formation through 
linear inertial wave propagation was observed. Large scales form columnar eddies aligned with 
the rotation axis and a linear growth of the axial integral length takes place once the Rossby number 
passes below a certain threshold ($R_o \sim 0.4$). 
With the increase of rotation, energy is retained by stable large scale structures and 
prevented form cascading to small scales. In some cases, energy was observed to decay faster for larger 
rotation frecuency.

In simulations, \cite{Thiele 2009, Muller 2007} reported depletion of the nonlinear energy 
cascade and growth of anisotropy. Also, an increase on the energy decay rate with rotation 
frequency was observed for the isotropic as well as the perpendicular modes. Two-dimensionalization 
was reported in several works (see e.g., \cite{Morinishib 2001}), together 
with the formation of columnar structures \cite{Kuczaj 2009} as seen in experiments. In 
\cite{Yang 2004}, three distinct regimes were observed depending on the rotation frecuency. 
At low rotation rates the flow behaves as non-rotating. At intermediate rotation 
rates, a strong coupling between rotation and non-linear interactions dominates (with a slower decay of the energy), 
and at high rotation rates viscous effects are dominant, damping the nonlinear effects. Recently, the 
cases of helical and non-helical rotating decaying turbulence with the integral scale of the 
size of the box were numerically studied in \cite{Teitelbaum 2009}, where it was found that 
the presence of net helicity decreases even further the decay rate of energy 
(see also \cite{Morinishi 2001}). Rotating helicoidal flows have applications in 
atmospheric research, helical convective storms being an example \cite{Lilly 1986}.

To explain some of the experimental and numerical results, an extension to 
phenomenological predictions for rotating turbulence for $E(k\rightarrow0) \sim Sk^2$ and 
$E(k\rightarrow0) \sim Ik^4$ initial spectra 
(usually known as Saffman and Batchelor spectra respectively) 
has been proposed \cite{Squires 1994}. It includes 
a slow-down factor of the energy flux due to the presence of Rossby waves involving two different 
timescales: a long timescale representative of the turbulence evolution and 
a short one associated with the rotation frequency $\Omega$. Conservation of 
$S$ and $I$ is then called to derive the asymptotic decay of energy for both initial 
spectra resulting in $E \sim t^{-3/5}$ and $E \sim t^{-5/7}$ respectively.
In the case of constrained turbulence, phenomenology leads to $E \sim t^{-1}$.
However, these phenomenological arguments do not consider the effect of anisotropies in the integral 
conserved quantities or in the decay laws.

In this work we numerically study the decay of rotating helical and 
non-helical turbulence with an emphasis on the anisotropies that arise when 
rotation is present, and on how integral quantities may be modified. The paper 
is divided in five sections. In section 2 we introduce the equations 
and describe how we solved them, with information regarding initial conditions.
In section 3 we consider as an example a non-rotating flow, which behaves in 
agreement with previous results.
In sections 4 and 5 we show and analyze results for the rotating non-helical 
and helical cases respectively. In the presence of rotation, we observe a decoupling of 
the energy decay rates for the 2D and three-dimensional (3D) modes. Studying the 
low wave number behavior we propose that the conservation of two-dimensional integral moments 
may explain these decays. In section 6 we finally summarize the results.

\section{Numerical Simulations}

The Navier-Stokes equation for an incompressible fluid in a rotating frame 
is solved numerically. When rotation is present, the equation reads

\begin{equation}
\partial_t {\bf u} + \mbox{\boldmath $\omega$} \times
    {\bf u} + 2 \mbox{\boldmath $\Omega$} \times {\bf u}  =
    - \nabla {\cal P} + \nu \nabla^2 {\bf u} ,
\label{eq:momentum}
\end{equation}
together with the incompressibility condition

\begin{equation}
\nabla \cdot {\bf u} =0.
\label{eq:incompressible}
\end{equation}
Here ${\bf u}$ is the velocity field, 
$\mbox{\boldmath $\omega$} = \nabla \times {\bf u}$ is the vorticity, 
${\cal P}=p/\rho-({\bf \Omega} \times {\bf r})^2/2+{\bf u}^2/2$ is the total modified 
pressure, $\rho$ is the (unit) density, and $\nu$ is the kinematic viscosity. We chose the rotation axis 
in the $z$ direction so that $\mbox{\boldmath $\Omega$} = \Omega \hat{z}$, 
$\Omega$ being the rotation frequency. Our integration domain is a cubic box of length 
$2\pi$ with periodic boundary conditions and the equations were solved using a 
pseudo-spectral method with the $2/3$-rule for de-aliasing.
All runs were performed with a resolution of $512^3$ grid points. 
The initial conditions were a superposition of Fourier modes with random phases 
and an energy spectrum $E(k) \sim k^4$ with modes randomly distributed in a spherical 
shell of wave numbers between 1 and 14. 
Details of the simulations are given in Table 1.

%%%%%%%%%%%%%%%%%[table with the runs information]%%%%%%%%%%%%%%%%%

\begin{center}
\tabcolsep=3pt  %%% This parameter is used to adjust the width of the table
\small
\renewcommand\arraystretch{1.2}  %% This command sets the vertical space between lines of the table
\begin{minipage}{15.5cm}{
\small{\bf Table 1.} Parameters used in the simulations: $t^*$ refers to the 
time of maximum enstrophy; $Re$, $R_o$, $R_o^w$, $H$ and $h$ are respectively 
the Reynolds, Rossby, micro-Rossby numbers, the total helicity and the relative 
helicity. All quantities are calculated at time $t^*$. A resolution of $512^3$ 
grid points was used in all runs.}
\end{minipage}
\vglue5pt

\begin{tabular}{| c | c | c | c | c | l | r | c | c |}  %% format a table with six rows
\hline %%
 {Run} & {$\nu$} & {$\Omega$} & {$Re$} & {$R_o$} & {$R_o^w$} & {$H$} & {$h$}& {$t^*$}\\   %% first line of the table
 \hline
  {1} & {$8.5\times10^{-4}$} & {$0$} & {$420$} & {$-$} & {$-$} & {$0.01$} & {$1\times 10^{-4}$} & {$0.6$}\\   %% second line of the table
  {2} & {$8.5\times10^{-4}$} & {$10$} & {$450$} & {$0.1$} & {$0.95$} & {$0.05$} & {$4\times 10^{-3}$} & {$0.7$}\\   %% third line of the table
  {3} & {$8.0\times10^{-4}$} & {$10$} & {$530$} & {$0.07$} & {$0.7$} & {$6.5$} & {$0.5$} & {$1.5$}\\   %% third line of the table
\hline
\end{tabular}
\end{center}

\vspace*{2mm}

%%%%%%%%%%%%%%%%%%%[en of table]%%%%%%%%%%%%%%%%%%%%%%%%%%%%%%%%%%%

In table 1, $Re=LU/\nu$ is the Reynolds number, $R_o^L=U/(2L\Omega)$ is the Rossby number based 
on the integral scale $L$, $R_o^w=w/(2\Omega)$ is the micro-Rossby number with 
$\omega=\langle {\mbox{\boldmath $\omega$}}^2 \rangle^{1/2}$, $H=\langle {\bf u \cdot \mbox{\boldmath $\omega$}} \rangle$ is the total helicity, 
$h=H/\langle|{\bf u}||\mbox{\boldmath $\omega$}|\rangle$ is the relative helicity, and $t^*$ 
is the time when maximum enstrophy is attained and when theses quantities are measured. We use $L$ 
defined as

\begin{equation}
L = 2\pi E^{-1}\int{E(k) k^{-1} dk}.
\end{equation}

Note that $R_o^w$ and $R_o^L$ are one order of magnitude apart. This is required for rotation 
not to completely damp the non-linear term in the Navier-Stokes equation leading to a pure 
exponential decay (see \cite{Cambon 1997} for more details).

\section{Non-rotating flow}
\subsection{Phenomenological arguments}

The classical Kolmogorov phenomenology 
leads to the well known energy spectrum
 
\begin{equation}
E(k)\sim \epsilon^{2/3}k^{-5/3}, 
\end{equation}
which for a decaying self-similar flow with $E(t)\sim kE(k)$ and using the balance 
equation $dE/dt \sim \epsilon$, gives the result

\begin{equation}
\frac{dE}{dt} \sim \frac{E^{3/2}}{L}. 
\end{equation}

$L$ can depend on time and extra hypothesis are required to obtain the 
energy decay. 

If $L \sim L_0$ (where $L_0$ is the size of the simulation domain), then $dE/dt \sim E^{3/2}/L_0$ 
and it follows that

\begin{equation}
E(t)\sim t^{-2}.
\end{equation}

\begin{figure}
\begin{center}
\includegraphics[width=12cm]{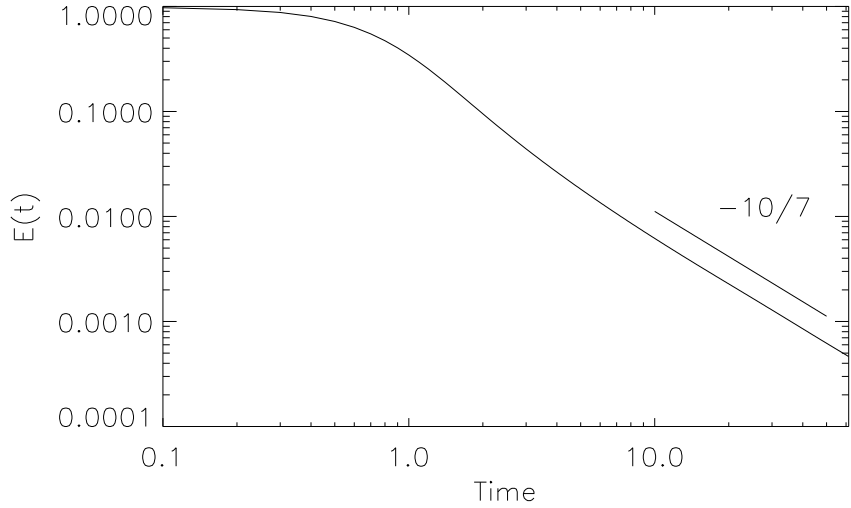}
\caption{Energy decay for the non-rotating case (run 1). After a transient, the self-similar 
decay agrees with classical Kolmogorov theory. The $t^{-10/7}$ slope 
is shown as a reference.}
\label{fig:decay_iso}
\end{center}
\end{figure}

If $L \neq L_0$ but the spectrum at large scales is $\sim k^4$, conservation of an integral 
quantity can be assumed to derive the decay rate of this flow. Traditionally the conservation 
of an initial $k^4$ dependence in the low wave number spectrum 
$E(k)$ has been related to the invariance of the Loitsyanski integral $I$ which we define as:

\begin{equation}
I = \int_{0}^{\infty}{r^4\langle {\bf u \cdot u'} \rangle dr},
\label{eq:I}
\end{equation}
where $\langle {\bf u \cdot u'} \rangle$ is the isotropic two-point longitudinal correlation function which depends solely on $r$. 
If conserved, from dimensional analysis it follows that $I \sim L^5U^2$, then 
$dE/dt \sim E^{17/10}/I^{1/5}$ and we finally get

\begin{equation}
E(t)\sim t^{-10/7}
\label{eq:Eiso}
\end{equation}
as obtained by Kolmogorov \cite{Kolmogorov 1941}.

In practice, $I$ evolves slowly in time and is only approximately conserved for an
initial large scale spectrum $\sim k^4$. If the large scale spectrum is $\sim k^2$, then another integral quantity is 
approximately conserved \cite{Saffman 1967}, which leads to a decay $E(t)\sim t^{-6/5}$. In the 
next section we present a simulation (run 1 of table 1) which approximately follows the 
Kolmogorov decay (see also \cite{Ishida 2006}). The rotating cases in section IV and V have the same 
large scale energy spectrum. Conditions where the integral length saturates  (reaching the box size 
in numerical simulations) have been reported in 
\cite{Teitelbaum 2009}.

\subsection{Numerical results}

In run 1 the energy spectrum (not shown) peaks initially at $k=14$ and 
maintains an approximately $k^4$ scaling for low wave-numbers. The time 
history $E(t)$ for run 1 is shown in figure \ref{fig:decay_iso}. 
After an initial transient of about six turn-over times, it shows a 
self-similar decay that is consistent with the $t^{-10/7}$ law.

In order to test further the behavior at large scales, we calculate the 
Loitsyansky integral $I$ for this isotropic flow. In a recent work
Ishida {\it et al.} estimated $I$ by fitting $E = Ik^4/24 \pi^2$ 
to the energy spectrum at large scales \cite{Ishida 2006}.
In their simulations (with spatial resolution of 1024 grid points and an 
initial peak of the spectrum near $k=40$ or $k=80$) the interval 
where $E \sim k^4$ holds is 
large enough for them to do the fitting. In our case (512 grid points) this interval 
is shorter and the fitting is not possible. Consequently, we 
checked spectral isotropy and then estimated $I$ using equation (\ref{eq:I}). 
The two-point longitudinal correlation function can be estimated in the isotropic case using \cite{Davidson 2004}:

\begin{equation}
\langle {\bf u.u'} \rangle(r) = 2\int_{0}^{\infty}{E(k)(sin kr - kr cos kr)/(kr)^3  dk},
\end{equation}
where $E(k)$ is the isotropic energy spectrum.

\begin{figure}
\begin{center}
\includegraphics[width=12cm]{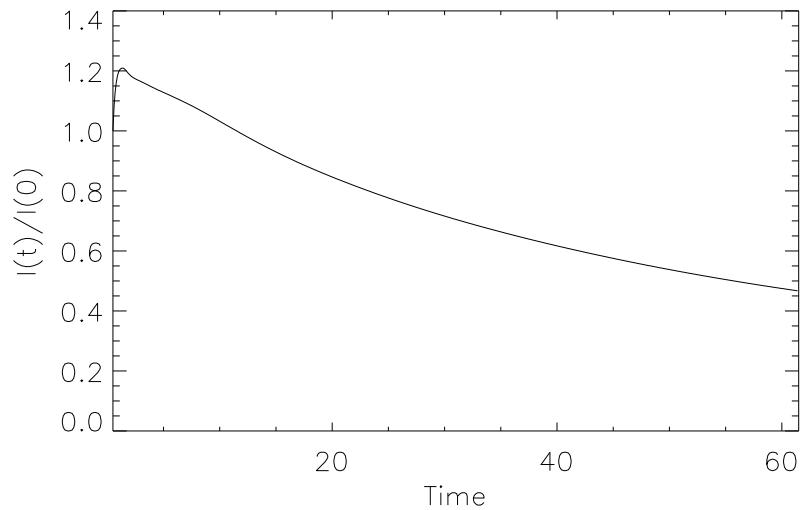}
\caption{Normalized Loitsyansky integral $I$ as a function of time for run 1.}
\label{fig:Ivst_3D}
\end{center}
\end{figure}

The evolution of $I(t)/I(0)$ is shown in figure \ref{fig:Ivst_3D}. 
Although $I(t)$ decays monotonically, after a transient its evolution is slow and it decreases only 
to approximately half its maximum value after 60 turn-over times. In \cite{Ishida 2006}
it was shown that its conservation improves as the extent of the large scale 
$\sim k^4$ spectrum is increased.

\begin{figure}
\begin{center}
\includegraphics[width=12cm]{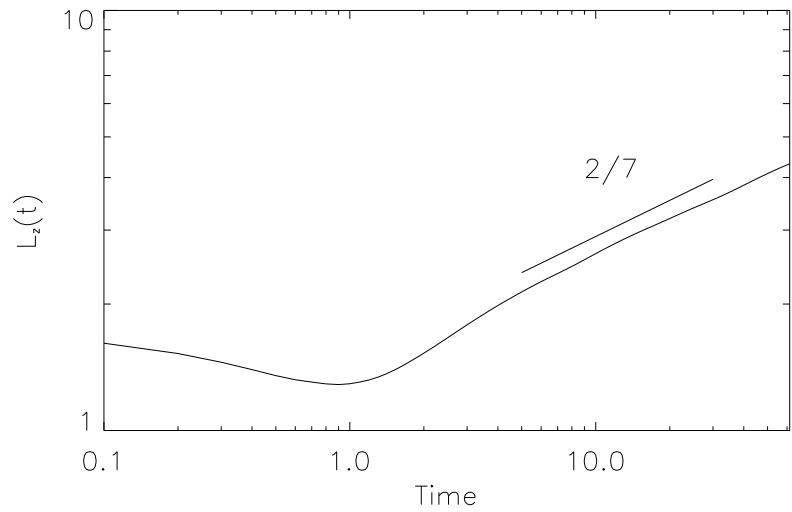}
\caption{One-dimensional integral length scale $L_z$ for run 1. After a transient 
period of time $L_z$ grows approximately as $t^{2/7}$ as derived phenomenologically.}
\label{fig:IntLpara}
\end{center}
\end{figure}

The approximate invariance of $I$ can further be used to estimate the growth 
of $L$. Writing $L \sim I^{1/5}U^{-2/5}$ an replacing in equation (\ref{eq:Eiso}) 
we get

\begin{equation}
L \sim t^{2/7}. 
\end{equation}
Figure \ref{fig:IntLpara} shows the evolution of the integral scale $L_z$ based 
on the one-dimensional spectrum for run 1 calculated as:

\begin{equation}
L_{z} = 2\pi E^{-1}\int{E(k_{z}) k_{z}^{-1} dk}.
\end{equation}
After an initial transient $L_z$ asymptotically settles down in the simulation 
to a growth rate close to $t^{2/7}$.

The results presented so far are consistent with Kolmogorov theory for 
decaying homogeneous and isotropic turbulence where initial conditions 
allow for the integral length to grow. In the next section we consider the 
analogy for the rotating case.

\section{Rotating flow}
\subsection{Phenomenological arguments}

In this section we analyze a flow subjected to solid-body rotation in the $z$
axis with rotation frequency $\Omega$ (run 2).
In non-helical rotating turbulence a 

\begin{equation}
E(k) \sim \epsilon^{1/2}\Omega^{1/2}k^{-2} 
\end{equation}
spectrum is typically assumed 
\cite{Muller 2007, Bellet 2006, Mininni 2009b, Zeman 1994, Zhou 1995}. 
From the balance equation this spectrum results in 

\begin{equation}
dE/dt \sim (E/L)^21/\Omega.
\label{eq:Eee}
\end{equation}

Again, there are at least three possible scenarios.
For $L \sim L_0$, $dE/dt \sim E^2$ \cite{Teitelbaum 2009} and 

\begin{equation}
E(t)\sim t^{-1}. 
\end{equation}

\begin{figure}
\begin{center}
\includegraphics[width=12cm]{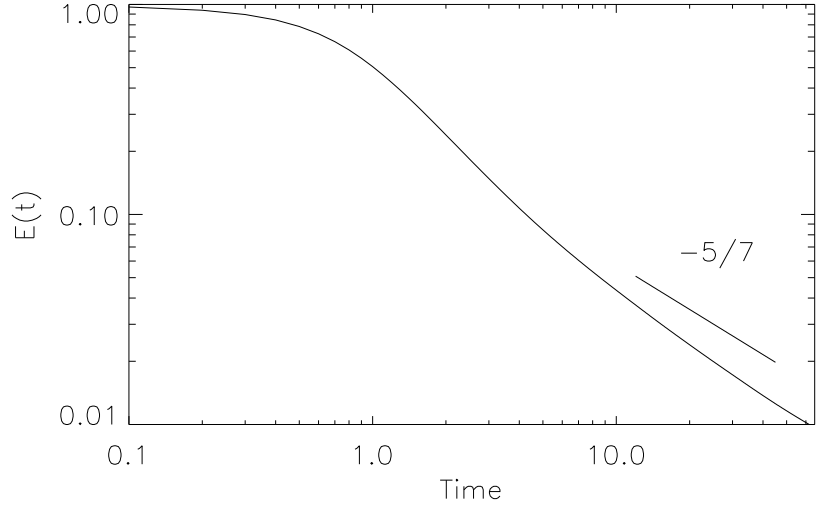}
\caption{Energy as a function of time for run 2. After a transient, a decay close to 
although shallower than $t^{-5/7}$ is attained. A $t^{-5/7}$ slope is 
plotted for reference.}
\label{fig:isodecayroth} 
\end{center}
\end{figure}

When $L \neq L_0$ and $E(k) \sim k^4$ at large scales as in our runs, 
constancy of $I\sim U^2L^5$ can be used again so that 
$dE/dt \sim E^{12/5}/(I^{2/5}\Omega)$ and \cite{Squires 1994}

\begin{equation}
E(t) \sim t^{-5/7}.
\end{equation}
Invariance of $I$ also leads to

\begin{equation}
L \sim t^{1/7}.
\end{equation}

Finally, details of the rotating case with $E(k)\sim k^2$ can be found in 
\cite{Squires 1994}.

\subsection{Numerical results}

Figure \ref{fig:isodecayroth} shows the evolution of $E(t)$. 
After an initial nearly inviscid period, a transient period leads 
to a decay rate slightly steeper than $E \sim t^{-5/7}$. 

\begin{figure}
\begin{center}
\includegraphics[width=12cm]{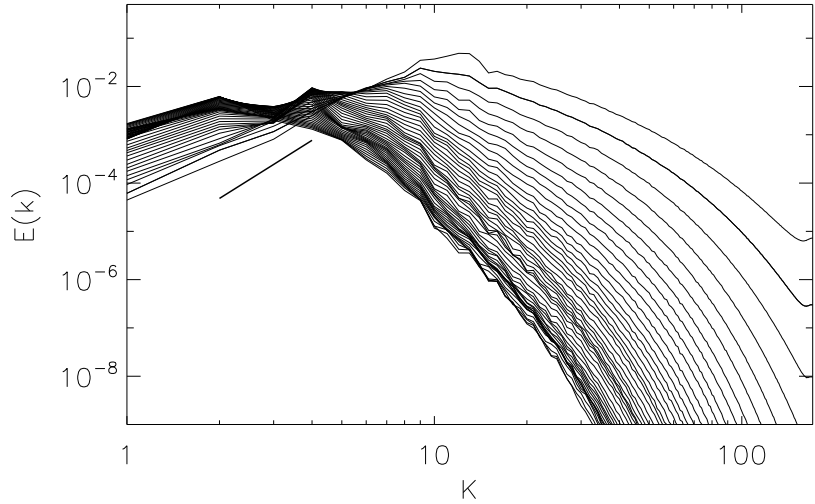}
\caption{Energy spectrum E(k,t) for run 2 from $t=1$ to $t=45$ in steps 
of $\Delta t = 1$. Energy piles up at large scales shallowing 
the spectra. A $k^4$ slope is shown for reference only.}
\label{fig:espectros3D_roth} 
\end{center}
\end{figure}

\begin{figure}
\begin{center}
\includegraphics[width=12cm]{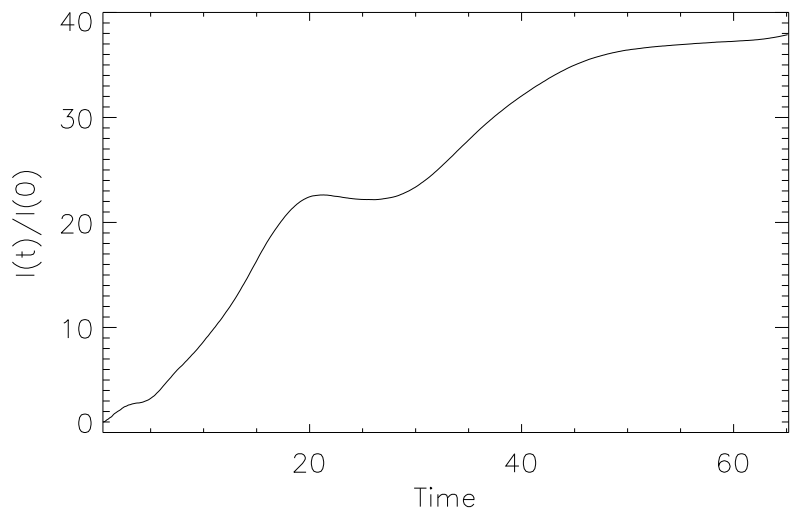}
\caption{Isotropic Loitsyansky integral $I$ for run 2}
\label{fig:Ivst_3D_roth} 
\end{center}
\end{figure}

The conservation of $I$ assumed to derive $E(t)\sim t^{-5/7}$ is associated to 
a preserved $k^4$ spectrum at large scales. However, in the presence of  rotation 
an inverse cascade of energy develops shallowing the spectrum at large scales 
as the energy piles up at low wave-numbers. 
This effect is visible in figure \ref{fig:espectros3D_roth} where the 
isotropic energy spectra is plotted for different times. As a consequence $I$ is no longer approximately conserved and varies fast. Figure \ref{fig:Ivst_3D_roth} 
shows the evolution of $I(t)/I(0)$ departing from a constant value and increasing by an order of magnitude in clear 
contradiction with the hypothesis of constant $I$.

\begin{figure}
\begin{center}
\includegraphics[width=12cm]{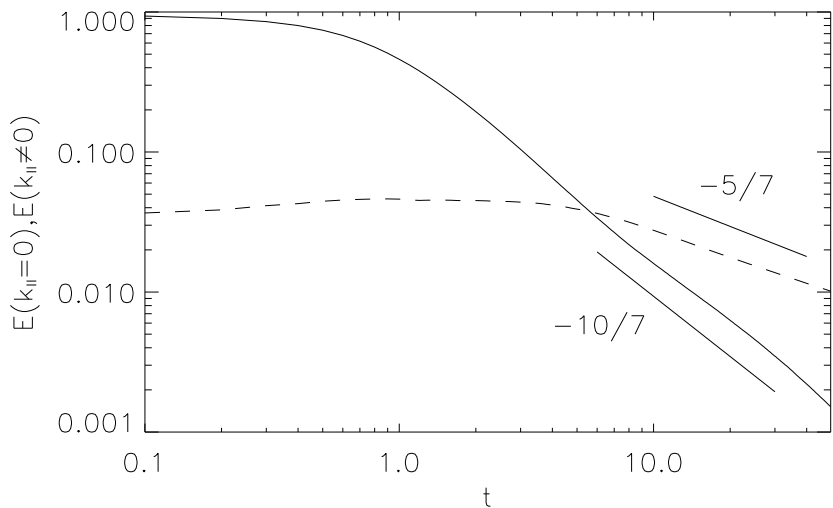}
\caption{$E(t)$ for 3D modes with $k_{\parallel} \neq 0$ (solid) and for 
2D modes in the $k_{\parallel}=0$ plane (dashed) for run 2. Note an apparent decoupling 
in the decay rate between the energy in the 2D and the 3D modes.}
\label{fig:decay_roth} 
\end{center}
\end{figure}

When rotation is present, isotropy breaks down, the flow becomes axisymmetric 
and a privileged direction (the $z$ axis) exists.
In figure \ref{fig:decay_roth} we show the evolution of the energy in the slow 2D modes 
of the velocity $E(k_{\parallel} =0,t)$ (where $k_{\parallel}=k_z$ are the 
wave numbers in the direction parallel to the rotation axis), together with the energy of the remaining 3D 
modes with $E(k_{\parallel} \neq 0,t)$. The $k_{\parallel} \neq 0$ modes dominate initially having 
an amplitude one order of magnitude larger than the modes with 
$k_{\parallel}=0$. 
This initial dominance is a result of the choice of initial conditions where Fourier modes 
are excited randomly all over shells in Fourier space, which results in most of 
the energy in modes with $k_{\parallel} \neq 0$.

As time evolves, energy from 3D modes is transferred to the 
$k_{\parallel} = 0$ 
plane and, after approximately six turn-over times, there is a cross-over after 
which energy in the 2D modes prevails. Thereafter, 3D modes decay as in an isotropic flow 
without rotation following $E(k_{\parallel} \neq 0) \sim t^{-10/7}$ while perpendicular modes 
follow a shallower decay close to but now shallower than $E_{\perp} \sim t^{-5/7}$. 
This suggests that after $t \approx 6$ the evolution of the 2D and 3D modes 
decouple and they decay separately with their own decay rates. The Rossby 
Number at that time is $R_o \approx 0.015$, and the behavior is 
consistent with predictions of wave turbulence theory that obtains a decoupling for 
very small Rossby number \cite{Majda 1998, Chen 2004} with the 2D evolution of the modes 
described by the 2D Navier-Stokes equation.   

\subsection{Phenomenology revisited}

This behavior naturally leads us to review two-dimensional integral moments 
in addition to the isotropic Loitsyansky integral already introduced. 
For 
two-dimensional turbulence, \cite{Davidson 2007} and \cite{Fox 2008} suggest 
that three canonical cases exist: 
$E(k\rightarrow0) \sim Jk^{-1}$, $E(k\rightarrow0) \sim Kk$ and 
$E(k\rightarrow0) \sim I_{2D}k^3$, where $J$, $K$ and $I_{2D}$ will be respectively defined here as

\begin{equation}
J=\int{\langle {\bf w.w'} \rangle r dr},
\end{equation}

\begin{equation}
K=\int{\langle {\bf u.u'} \rangle r dr}, 
\end{equation}
and 

\begin{equation}
I_{2D}=\int{r^3\langle {\bf u.u'} \rangle dr}.
\end{equation}

Moreover, $J$ and $K$ are integral invariants of motion with conservation 
of $K$ being associated with conservation of linear momentum and invariance 
of $J$ a consequence of vorticity conservation. In the 3D rotating case, 
since the $k_{\parallel}=0$ modes are the ones approximately decoupled for $R_o \ll 1$, and 
the equations for these modes are equivalent to the 2D Navier-Stokes equations 
\cite{Majda 1998, Chen 2004}, we may wonder whether these integral quantities are 
conserved (or at least evolve slowly with time) in that manifold. In the rotating case, the relevant increments 
are then $r=r_{\perp}$ with ${\bf r_{\perp}}$ perpendicular 
to ${\bf \Omega}$, and the associated wave vectors are ${\bf k_{\perp}}$.

\begin{figure}
\begin{center}
\includegraphics[width=12cm]{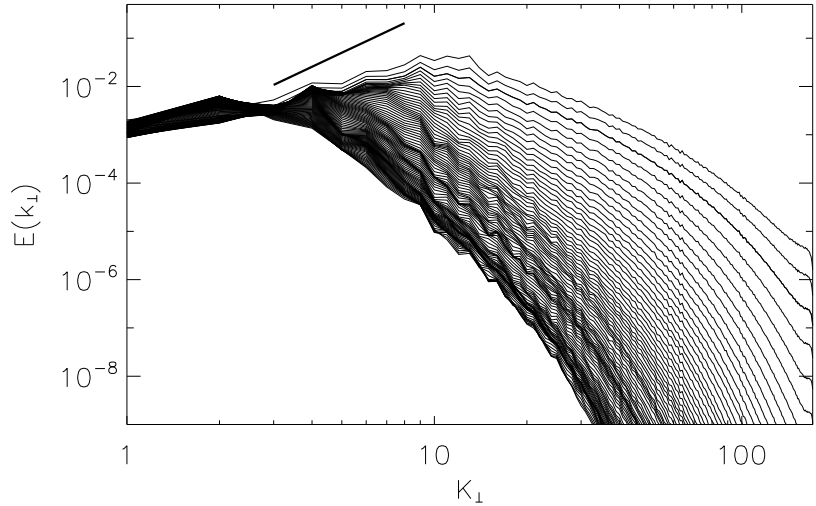}
\caption{Evolution of the two-dimensional perpendicular spectrum $E(k_{\perp},t)$ for run 2 from 
$t=0.5$ to $t=45$ in steps of $\Delta t=0.5$.
Note a slightly shallower than $k^3$ scaling for low wave numbers.}
\label{fig:espectros2D} 
\end{center}
\end{figure}

We start by showing in figure \ref{fig:espectros2D} the energy spectrum for the vertically-averaged 
velocity field, that is to say, for wave numbers $k_{\perp}^2=k_x^2 + k_y^2$ 
(hereafter, the perpendicular spectrum). This spectrum maintains 
a form proportional to $k_{\perp}^3$ (albeit slightly shallower).

\begin{figure}
\begin{center}
\includegraphics[width=12cm]{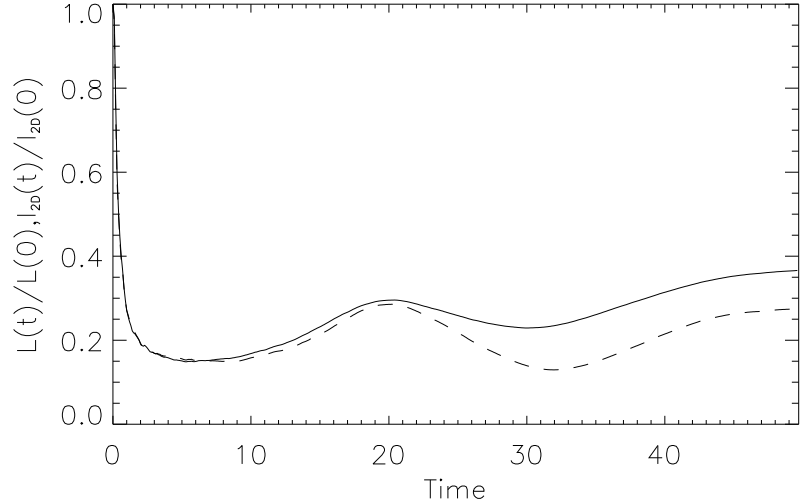}
\caption{Normalized $K$ (solid) and $I_{2D}$ (dashed) as a function of time for run 2. 
Both parameters have a slow, almost constant behavior in time, showing little variation.}
\label{fig:Lvst_2D} 
\end{center}
\end{figure}

In order to find slowly varying 2D-like integral quantities in the 
simulation, we calculate the time 
evolution of $K$ and $I_{2D}$ inverting the following equation 
which follows from assuming axisymmetry \cite{Davidson 2004}:

\begin{equation}
E(k_{\perp})= \int{\frac{1}{2} \langle {\bf u.u'}\rangle k_{\perp}r_{\perp}J_0(k_{\perp}r_{\perp}) dr},
\label{eq:tom}
\end{equation}
so that

\begin{equation}
\langle {\bf u.u'}\rangle(r_{\perp}) = \int{2E(k_{\perp})J_0(k_{\perp}r_{\perp})dk_{\perp}}
\end{equation}
can be estimated from the perpendicular spectrum.
The results for $K(t)/K(0)$ and $I_{2D}(t)/I_{2D}(0)$ are plotted in figure \ref{fig:Lvst_2D}. 
Both magnitudes behave in a similar fashion, showing slow variations with a relative constant value over the simulated time.

Invariance of $K$ or $I_{2D}$ leads to different energy decay rates. 
For constant $K$ we can write $K \sim L_{\perp}^2U_{\perp}^2L_{0\parallel}$ 
(where $L_{0\parallel}$ is the size of the box in the direction parallel to ${\mbox{\boldmath $\Omega$}}$) and assuming the slow-down 
factor in the energy dissipation rate by waves, as done in equation (\ref{eq:Eee}), 
$dE_{\perp}/dt \sim (E_{\perp}/L_{\perp})^2/\Omega$. Replacing $L_{\perp}$ in the last 
equation we get $dE_{\perp}/dt \sim E_{\perp}^3L_{0 \parallel}/(K\Omega)$ leading to

\begin{equation} 
E_{\perp} \sim t^{-1/2}.
\end{equation}

For constancy of $I_{2D}$ we have 
$I_{2D} \sim L_{\perp}^4U_{\perp}^2L_{0\parallel}$, and using the same arguments, 
from $dE_{\perp}/dt \sim (E_{\perp}/L_{\perp})^2/\Omega$ 
we get $dE_{\perp}/dt \sim E_{\perp}^{5/2}L_{0 \parallel}^{1/2}/(I_{2D}^{1/2}\Omega)$, which finally leads to 

\begin{equation}
E_{\perp} \sim t^{-2/3}.
\end{equation}

\begin{figure}
\begin{center}
\includegraphics[width=12cm]{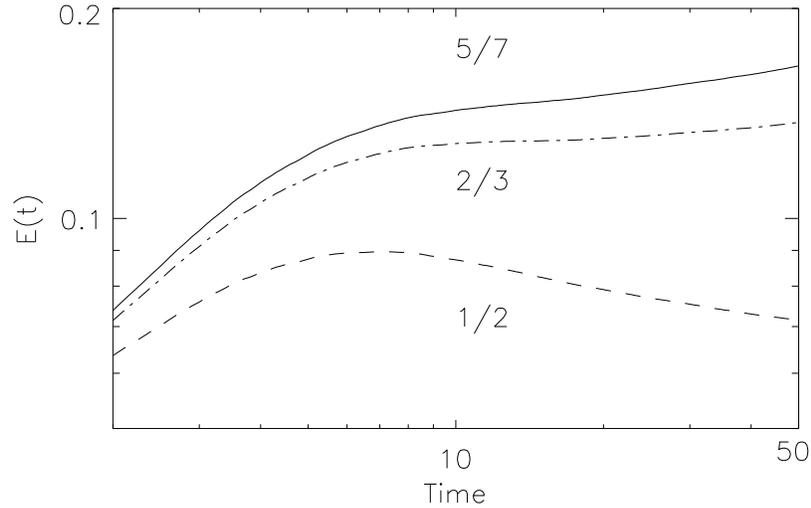}
\caption{Energy decay for perpendicular ($k_{\parallel}=0$) modes compensated by 
$t^{-\alpha}$ for $\alpha=5/7$ (solid), $\alpha=2/3$ (dot-dashed) and 
$\alpha=1/2$ (dashed); $\alpha=2/3$ adjusts our data better.}
\label{fig:compensated} 
\end{center}
\end{figure}

In order to see whether any of these decay laws adjust our data better than 
the isotropic $\sim t^{-5/7}$ law, we plot the 2D
energy evolution compensated by $t^{-5/7}$, $t^{-1/2}$, and $t^{-2/3}$ (figure \ref{fig:compensated}). The decay law $E_{\perp} \sim t^{-2/3}$ seems to adjust better 
our simulation, which is consistent with the slow variation of $I_{2D}$ 
and with the initial perpendicular spectrum close to $k_{\perp}^{3}$.

\begin{figure}
\begin{center}
\includegraphics[width=12cm]{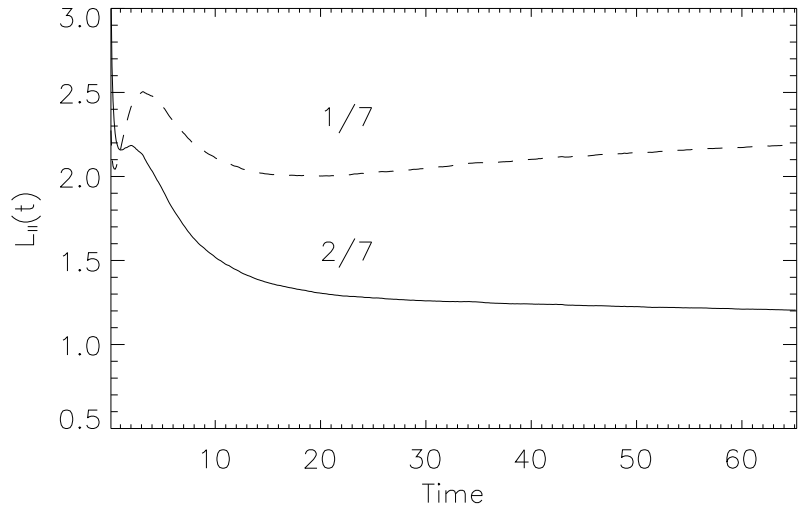}
\caption{Integral length parallel to the rotation axis for run 2 compensated by 
$t^{-\alpha}$ for $\alpha=2/7$ (solid) and $\alpha=1/7$ (dashed); $t^{2/7}$ gives a 
better agreement to our data.}
\label{fig:Lpara_roth_compensado} 
\end{center}
\end{figure}

The integral length parallel to the rotation axis $L_{\parallel}=L_z$ also behaves 
as in the isotropic case. In figure \ref{fig:Lpara_roth_compensado} we plot the 
evolution of $L_{\parallel}$ compensated by the laws for the isotropic and axisymmetric cases already 
deduced. Clearly, the isotropic $t^{2/7}$ scaling adjusts better the 
results, in agreement with the isotropic-like decay of the energy in the 3D 
fast modes.

\section{Helical Rotating Flow}

We finally discuss briefly the effect of helicity upon the decay rate of energy in a rotating fluid. In order to incorporate net helicity into the flow we use  a superposition of 
Arnold-Beltrami-Childress (ABC) \cite{Childress 1995} initial conditions 
achieving an initial relative helicity $h\approx 0.99$. As in run 2, the ABC 
flows were added in all shells in Fourier space between wave numbers $k=1$ to 
$14$ with an isotropic spectrum $\sim k^4$.

\begin{figure}
\begin{center}
\includegraphics[width=12cm]{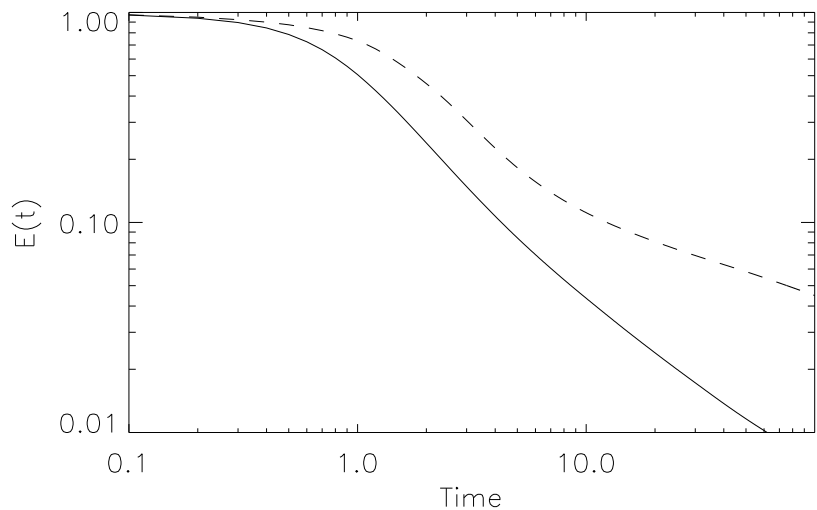}
\caption{Total isotropic energy decay for helical (dashed) and non-helical (solid) rotating turbulence.
The results show a decreased decay rate in the presence of helicity.}
\label{fig:decay_roth_heli1} 
\end{center}
\end{figure}

Figure \ref{fig:decay_roth_heli1} shows a comparison between the energy as a 
function of time for the helical and non-helical rotating flows. 
The helical energy decay, shown in dashed line, is 
slower than the non-helical case. This retard has been associated with an inhibition 
of the non-linear transfer of the energy toward smaller scales by a direct cascade of helicity \cite{Mininni 2009}.  
The behavior has also been observed in \cite{Teitelbaum 2009}, where rotating 
helical and non-helical flows with constant integral length were studied.

The distinct evolution in the free decay of the helical flow can also 
be understood in terms of 
a phenomenological theory similar to those already presented. In this case, the direct 
transfer is dominated by the helicity cascade. Assuming maximal helicity we 
have \cite{Mininni 2009} 

\begin{equation}
E(k_{\perp}) \sim \epsilon^{1/4}\Omega^{5/4}k_{\perp}^{-5/2}.
\end{equation}
Then it follows that  

\begin{equation}
dE_{\perp}/dt \sim E_{\perp}^4/(L_{\perp}^6\Omega^5).
\end{equation}

\begin{figure}
\begin{center}
\includegraphics[width=12cm]{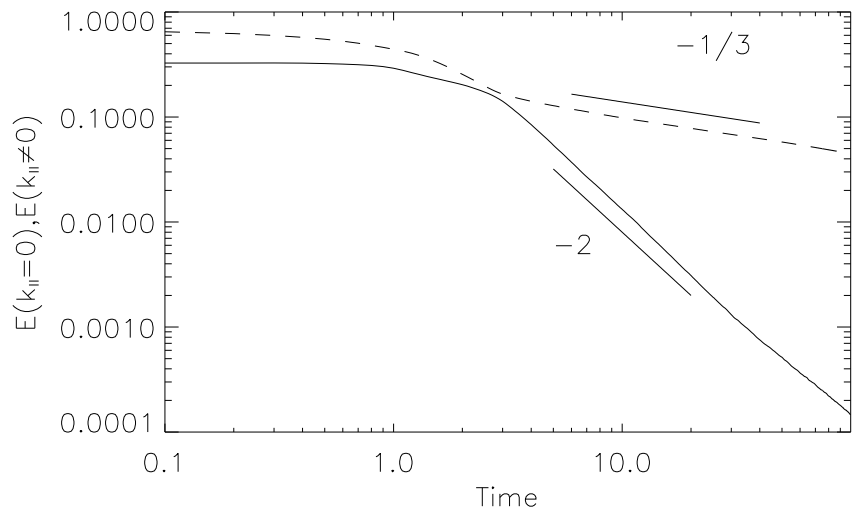}
\caption{Decay of the energy for 3D modes with $k_{\parallel} \neq 0$ (solid) and for 
2D modes in the $k_{\parallel}=0$ plane (dashed) for the helical rotating flow (run 3) showing 
different scaling laws.}
\label{fig:decay_roth_heli} 
\end{center}
\end{figure}

\begin{figure}
\begin{center}
\includegraphics[width=12cm]{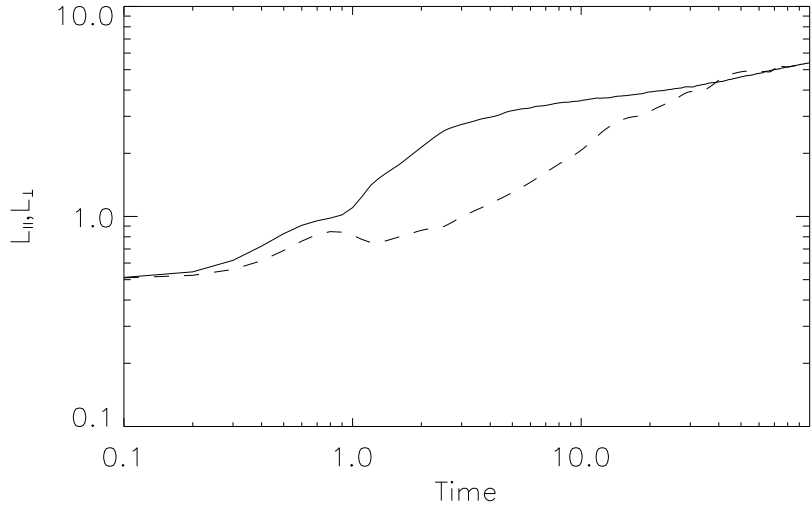}
\caption{Integral length scales parallel (solid) and perpendicular (dashed) 
to the rotation axis as a function of time. Both lengths saturate
fast to an almost constant value near the simulation domain length 
($2\pi$).}
\label{fig:Lrothheli} 
\end{center}
\end{figure}

\begin{figure}
\begin{center}
\includegraphics[width=12cm]{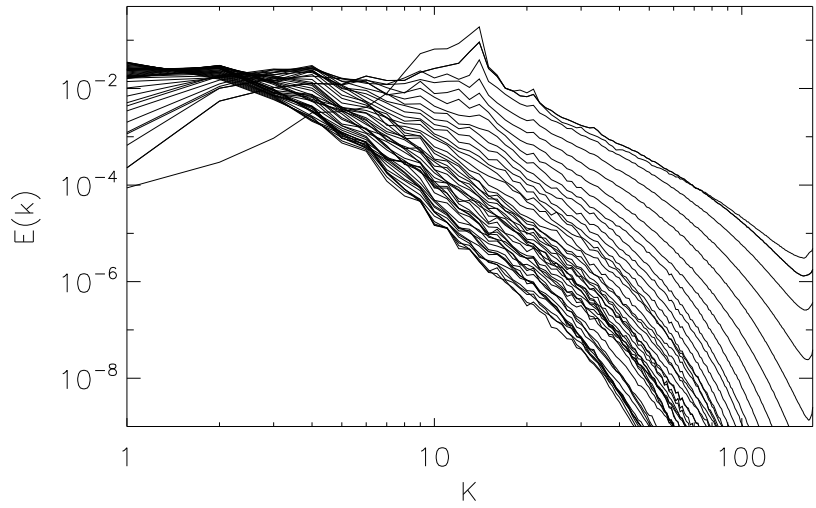}
\caption{Isotropic energy spectrum E(k,t) for run 3 from $t=1$ to $t=45$ in steps 
of $\Delta t = 1$.}
\label{fig:espectros3D_roth_heli} 
\end{center}
\end{figure}

It is unclear at this point whether in the phenomenological analysis we should separate 
the decay of the 2D modes from the 3D modes as in run 2, and whether the integral scale changes in 
time or not. Therefore, in figure \ref{fig:decay_roth_heli} we plot the 2D and 3D energy decays for run 3 
(same as figure \ref{fig:decay_roth} but for the helical case). Note the anisotropic 
initial state with a relative excess of energy in the modes with $k_{\parallel}=0$ due to the 
ABC initial conditions used. After the first nearly inviscid 
transient, both sets of modes seem again to decouple and decay with different laws: $\sim t^{-2}$ 
for the 3D modes and $\sim t^{-1/3}$ for the 2D modes. As in the non-helical case, the 3D modes decay 
faster, following the same law 
derived for an isotropic non-rotating flow where $L \sim L_0$. Indeed, in this run the 
integral scales grow fast during the transient and reach lengths close to the size of 
the box $L_0$ ($=2 \pi$) before the self-similar decay starts. This is illustrated in 
figure \ref{fig:Lrothheli}, that shows the evolution of the parallel and perpendicular integral scales. 
After $t\approx 6$ $L_{||}$ is almost saturated, and $L_\perp$ keeps growing slowly but close 
to its maximum value. This results from a fast inverse transfer of energy (see the evolution 
of the isotropic energy spectrum in figure \ref{fig:espectros3D_roth_heli}) that may be associated to the large amount of energy 
in the $k_{||}=0$ modes in the initial conditions.

The fast increase and saturation of $L_{||}$ and $L_\perp$ give as a result the decay of the 2D 
and 3D modes as in constrained turbulence. For the 3D modes the $\sim t^{-2}$ decay then follows. 
For the 2D modes, using equation (25) and the approximate constancy of the integral lengths, we get a decay

\begin{equation}
E_\perp \sim t^{-1/3}
\end{equation}
in agreement with the simulation. The study of the cases where the integral scales are not constant are left for future work, and may require the identification of anisotropic integral conserved quantities as in the previous section.

\begin{figure}
\begin{center}
\includegraphics[width=12cm]{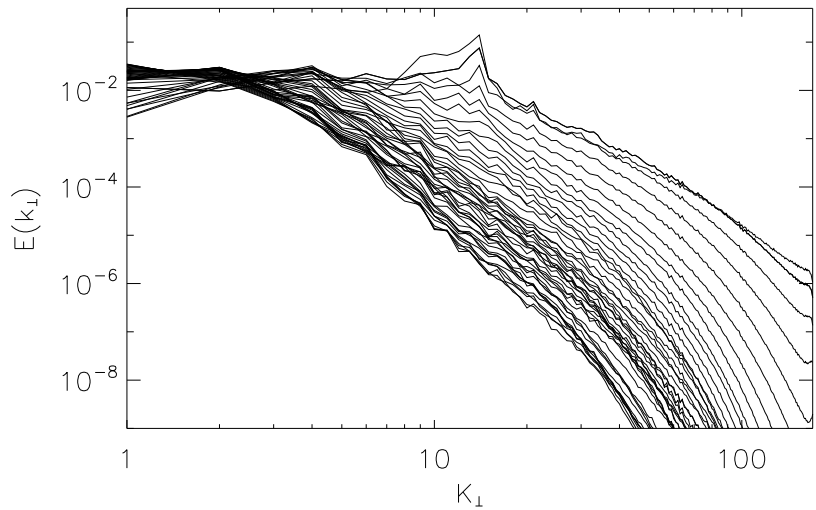}
\caption{Perpendicular energy spectrum $E(k_{\perp},t)$ for run 3 from $t=1$ to $t=45$ in steps 
of $\Delta t=1$.}
\label{fig:espectros3D_roth_heli_perp} 
\end{center}
\end{figure}

\section{Conclusion}

In this work we presented studies of decaying turbulence in the presence of rotation and helicity. A simulation (run 1) without rotation and helicity was used to introduce some of the phenomenological arguments in unconstrained decaying turbulence. The simulation, with a large scale spectrum close to $\sim k^4$, shows a slowly varying Loitsyansky integral and a decay law consistent with Kolmogorov's $E(t) \sim t^{-10/7}$ law. Detailed studies of such a decay can be found in \cite{Ishida 2006}.

When extending these arguments to rotating turbulence, approximate conservation of isotropic integral moments (as e.g., the Loitsyansky integral) is often assumed (see e.g., \cite{Squires 1994}). A simulation of non-helical rotating turbulence (run 2) was shown to decay slightly faster than what is predicted by these arguments. We argued that the approximate decoupling of slow and fast modes predicted in wave turbulence theory for rotating flows at very small Rossby numbers leads to different decay laws for the energy in the 2D and 3D modes. The 3D modes decay in agreement with phenomenological predictions for isotropic and homogeneous turbulence, while the decay of the 2D modes is consistent with phenomenological results obtained assuming integral moments of the two-dimensional Navier-Stokes equation are approximately conserved.

Finally, the effect of helicity in rotating turbulence was considered in run 3. Helicity decreases the decay rate of turbulence even further as the direct transfer is dominated by the direct helicity flux (see e.g., \cite{Mininni 2009, Teitelbaum 2009}), and helicity tends to decrease the amplitude of the non-linear term in the Navier-Stokes equation. The initial conditions considered led to the fast saturation of the integral scales, and as a result the 2D and 3D modes in this run decayed as constrained turbulence. The 3D modes decayed as in the non-rotating (constrained) case, while the 2D modes were observed to decay slower than what is predicted for constrained non-helical rotating turbulence and in agreement with predictions that consider the effect of helicity.

The three simulations presented here are far from an exhaustive exploration of the possible decay laws that may develop in rotating turbulence, and a detailed study of the effect of changing the initial large-scale energy spectrum dependence is left for future work, as well as studies of the effect of initial anisotropies in the decay, and the effect of scale separation between the initial integral scale and the box size (see e.g., \cite{Ishida 2006} for a study for isotropic and homogeneous turbulence), and parametric studies varying the Reynolds and the Rossby numbers.

\section*{Acknowledgments}
Computer time was provided by NCAR and CeCAR. The authors acknowledge support from grant 
UBACYT X468/08 and PICT 2007-02211. PDM is a member of the Carrera del Investigador Cient\'{\i}fico of CONICET.

%\bibliographystyle{acm}
%\bibliography{ms}

\section*{References}

\begin{thebibliography}{10}

\bibitem{Davidson 2004}
{Davidson P A 2004 {\it Turbulence: An introduction for scientists and engineers} 
(Oxford University Press)}

\bibitem{Kolmogorov 1941}
{Kolmogorov A N 1941 On the degeneration of isotropic turbulence in an 
incompressible viscous fluid {\it Dokl. Akad. Nauk SSSR} {\bf 31} 538-541}

\bibitem{Saffman 1967}
{P G Saffman 1967 Note on decay of homogeneous turbulence {\it Phys.
Fluids} {\bf 10} 1349}

\bibitem{Herring 2005}
{Herring J R, Kimura Y, James R, Clyne J, and Davidson P A 2005 Statistical and 
dynamical questions in stratified turbulence. In {\it Mathematical and Physical Theory of 
Turbulence} (ed. S. Shivamoggi) Taylor Francis}

\bibitem{Ishida 2006}
{Ishida T, Davidson P A, and Kaneda Y 2006 On the decay of isotropic turbulence 
{\it J.Fluid Mech.} {\bf 564} 455-475}

\bibitem{Davidson 2007}
{Davidson P A 2007 On the large-scale structure of homogeneous two-dimensional 
turbulence {\it J.Fluid Mech.} {\bf 580} 431-450}

\bibitem{Mansour 1992}
{Mansour N N, Cambon C, and Speziale C G 1992 Theoretical and computational study of rotating isotropic turbulence, in Studies in Turbulence, edited by T.B Gatski, S.Sarkar and C.G. Speziale, (Springer-Verlag)}

\bibitem{Squires 1994}
{Squires K D, Chasnov J R, Mansour N N, and Cambon C 1994 The asymptotic state of rotating homogeneous turbulence at high reynolds numbers, Application of direct and large eddy simulation to transition and turbulence" Chania, Crete, Greece}

\bibitem{Morize 2005}
{Morize C, Moisy F, and Rabaud M 2005 Decaying grid-generated turbulence in a 
rotating tank {\it Phys. Fluids} {\bf17} 095105-095105-11}

\bibitem{Morize 2006}
{Morize C and Moisy F 2006 Energy decay of rotating turbulence with confinement effects 
{\it Phys. Fluids} {\bf 18} 065107-065107-9}

\bibitem{Cambon 2004}
{Cambon C, Rubinstein R, and Godeferd F S 2004 Advances in wave turbulence: rapidly rotating flows
{\it New J. Phys.} {\bf 6} 73}

\bibitem{Bardina 1985}
{Bardina J, Ferzinger J H, and Rogallo R S 1985 Effect of rotation on isotropic 
turbulence: computation and modeling {\it J. Fluid Mech.} 
{\bf 154} 321-336}

\bibitem{Bartello 1994}
{Bartello P, Metais O, and Lesieur M 1994 Coherent structures in rotating 
three-dimensional turbulence {\it J. Fluid Mech.} {\bf 273} 1-29}

\bibitem{Greenspan 1968}
{Greenspan H P 1968 {\it The Theory of Rotating Fluids} (Cambridge University Press)}

\bibitem{Greenspan 1969}
{Greenspan H P 1969 On the nonlinear interaction of inertial waves {\it J. Fluid 
Mech.} {\bf 36} 257-286}

\bibitem{Waleffe 1993}
{Waleffe F 1993 Inertial transfers in the helical decomposition {\it Phys. 
Fluids} A {\bf 5} 677-685}

\bibitem{Cambon 1997}
{Cambon C, Mansour N N, and Godeferd F S 1997 Energy transfer in rotating 
turbulence {\it J.Fluid Mech.} {\bf 337} 303-332}

\bibitem{Chen 2004}
{Chen Q, Chen S, Eyink G L, and Holm D D 2005 Resonant 
interactions in rotating homogeneous three-dimensional turbulence {\it J.Fluid Mech.} {\bf 542} 139-164}

\bibitem{Majda 1998}
{Majda A J and Embid P F 1998 Averaging over fast gravity waves for geophysics flows 
with unbalanced initial data {\it Theor. Comp. Fluid Dyn.} {\bf 11} 155-169}

\bibitem{Seiwert 2008}
{Seiwert J, Morize C, and Moisy F 2008 On the decrease of intermittency in decaying 
rotating turbulence {\it Phys. Fluids} {\bf 20} 071702-071702-4}

\bibitem{Moisy 2009}
{Moisy F, Morize C, Rabaud M, and Sommera J 2009 Anisotropy and cyclone-anticyclone asymmetry in decaying 
rotating turbulence arXiv:0909.2599}

\bibitem{Staplehurst 2008}
{Staplehurst P J, Davidson P A, and Dalziel S B 2008 Structure formation in homogeneous freely decaying rotating turbulence
{\it J. Fluid Mech.} {\bf 598} 81-105}

\bibitem{Thiele 2009}
{Thiele M and Muller W C 2009 Structure and decay of rotating homogeneous turbulence arXiv:0906.0853}

\bibitem{Muller 2007}
{Muller W C and Thiele M 2007 Scaling and energy transfer in rotating 
turbulence {\it Europhy. Lett.} {\bf 77} 3 34003-34003-5}

\bibitem{Morinishib 2001}
{Morinishi Y, Nakabayashi K, and Ren S Q 2001 Dynamics of anisotropy on decaying homogeneous turbulence subjected to system rotation
{\it Phys. Fluids} {\bf 13} 2912-2922}

\bibitem{Kuczaj 2009}
{Kuczaj A K, Geurts B J, and Holm D D 2009 Intermittency effects in rotating decaying turbulence arXiv:0904.0713}

\bibitem{Yang 2004}
{Yang X, Domaradzki J A 2004 Large eddy simulations of decaying rotating turbulence 
{\it Phys. Fluids} {\bf 16} 4088-4104}

\bibitem{Teitelbaum 2009}
{Teitelbaum T and Mininni P D 2009 Effect of helicity and rotation on the free 
decay of turbulent flows {\it Phys.Rev.Lett.} {\bf 103} 014501}

\bibitem{Morinishi 2001}
{Morinishi Y, Nakabayashi K, and Ren S 2001 Effects of helicity and system rotation 
on decaying homogeneous turbulence 
{\it JSME Int. J. Ser. B} {\bf 44} 410-418} 

\bibitem{Lilly 1986}
{Lilly D K 1986 The structure, energetics and propagation of convective 
rotating storms. Part II:Helicity and storm stabilization {\it atm. sc.} {\bf 43} 126-140}

\bibitem{Bellet 2006}
{Bellet F, Godeferd F S, Scott J F, and Cambon C 2006 Wave tubulence in rapidly rotating flows {\it J.Fluid Mech.} {\bf 562} 83-121}

\bibitem{Mininni 2009b}
{Mininni P D, Alexakis A, and Pouquet A 2009 Scale interactions and scaling 
laws in rotating flows at moderate Rossby numbers and large 
Reynolds numbers {\it Phys. Fluids} {\bf 21} 015108}

\bibitem{Zeman 1994}
{Zeman O 1994 A note on the spectra and decay of rotating homogeneous turbulence {\it Phys. Fluids} {\bf 6} 10 3221-3223}

\bibitem{Zhou 1995}
{Zhou Y 1995 A phenomenological treatment of rotating turbulence {\it Phys. Fluids} {\bf 7} 2092-2094}

\bibitem{Fox 2008}
{Fox S and Davidson P A 2008 Integral invariants of two-dimensional and 
quasigeostrophic shallow-water turbulence {\it Phys. Fluids} {\bf 20} 075111}

\bibitem{Childress 1995}
{Childress S and Gilbert A D 1995 {\it Stretch, Twist, Fold: The fast dynamo} 
(Springer-Verlag Berlin)}

\bibitem{Batchelor 1956}
{Batchelor G K and Proudman I 1956 The large-scale structure of homogeneous 
turbulence {\it Phil. Trans. R. Soc. A} {\bf 248} 369-405}

\bibitem{Ossia 2000}
{Ossia S and Lesieur M 2000 Energy backscatter in large-eddy simulations of 
three-dimensional incompressible isotropic turbulence {\it J.Turbulence} 
{\bf 1} 10(1)}

\bibitem{Lesieur 2000}
{Lesieur M and Ossia S 2000 3D isotropic turbulence at very high Reynolds numbers: 
EDQNM study {\it J. Turbulence} {\bf 1}(1) 7(1)}
\bibitem{Mininni 2009}
{Mininni P D and Pouquet A 2009 Helicity cascades in rotating turbulence {\it Phys. Rev. E} {\bf 79} 026304}


\end{thebibliography}

\end{document}